\documentclass[showpacs,preprint2,amsmath,amssymb]{revtex4}
\usepackage{graphicx}
\usepackage{dcolumn}
\usepackage{subfigure}
\usepackage{epsfig}
\usepackage{bm}
\usepackage{caption}
\newcommand{\barr}{\begin{eqnarray}}
\newcommand{\earr}{\end{eqnarray}}

\newcommand\Tstrut{\rule{0pt}{2.6ex}}  
\newcommand\Bstrut{\rule[-1.2ex]{0pt}{0pt}}
\newcommand{\dis}{\displaystyle}

\usepackage{hyperref}

\begin{document}


\title{Supernova neutrino scattering
off Gadolinium odd isotopes in water Cherenkov detectors.}

\author{Paraskevi C. Divari}
\affiliation{Department of Physical Sciences and Applications,
Hellenic Military Academy, Vari 16673, Attica, Greece}

\begin{abstract}
  In this work  the supernova neutrino(SN)
charged-current interactions with     Gd odd isotopes (A=155 and
157) are studied.   We use measured spectra and the
quasiparticle-phonon model (MQPM)     to calculate the charged
current response of odd Gd isotopes to supernova neutrinos.
Flux-averaged cross sections are obtained considering
quasi-thermal neutrino spectra.
\end{abstract}
 \maketitle

\section{ Introduction}

Neutrino physics has become the main focus of astrophysics, particle
physics and nuclear physics. The Galaxy Supernova only happens once
every 30 years. Various detectors must be prepared to detect all
flavors of neutrinos well to understand the physics and astrophysics
of nuclear collapsed supernova. Neutrino oscillations provide solid
evidence for phenomena other than standard model physics, and the
experimental work used to detect weakly interacting particles is
more powerful than ever.
Using current detectors, it is easy to measure supernovae
$\bar\nu_e$ and $\nu_x$ via inverse beta decay (IBD) and elastic
scattering on protons, respectively
\cite{Vogel:1999zy,Strumia2003,Farr}. Detecting supernova neutrinos
is an indispensable method for testing supernova and neutrino
physics, but the yield is small, and the background from other
channels is also very large.


A number of detectors like water Cherenkov detectors (WCDs)
\cite{IMB,SK,SNO}, have been used in various neutrino detection
experiments. The neutrino detection in WCDs is based on the
detection of Cherenkov radiation. A charged particle moving
through at a speed greater than the speed of light in the media
will generate Cherenkov radiation.  The radiation is emitted at an
angle $\theta$
to the direction of travel, which are then
detected by the photomultiplier tubes  (PMTs) lined in the inner
walls of the detector. Neutrinos in water can be detected thanks to
several reactions. The most important  are the IBD, the elastic
scattering on electrons and the neutral current scattering on oxygen
\text{\textsuperscript{16}O}.

In an inverse beta decay, an electron antineutrino scattering from a
proton, creates a positron and a neutron. The positron undergoes a
prompt matter-antimatter annihilation emitting light   and the
neutron capture on hydrogen nucleus release not detectable 2.2 MeV
gamma cascades (energy threshold in Super-Kamiokande(SK) about 5
MeV). The sensitivity of the detectors can be enhanced including
additives, such as gadolinium-based salt compounds, in water that
essentially reduces background signals \cite{Laha}. The neutron
capture by gadolinium (Gd) results in 8 MeV gamma cascades with
temporal   and spatial   coincidence with the positron from the
initial interaction. This delayed neutron capture by Gd and higher
gamma emission energy compared to neutron capture by hydrogen
nucleus, creates a unique signature and improves background
reduction to an antineutrino event and thus allows antineutrino
detection in WCDs through IBD. In SK,  which is a 32~ktons
(fiducial) WCD,  it has be found that  the inclusion of
$\text{GdCl}{_3}$ salt (0.2\% in weight)  to SK, $\sim 90\%$ of the
IBD events could be tagged \cite{Laha,Vagins}. The remaining IBD
events  as well as the $\bar{\nu}_{e}$ absorption events on
\text{\textsuperscript{16}O} can then be statistically subtracted
from the remaining signal.

In Ref.~\cite{Divari2018} we payed  special attention on
calculations of charged current (CC) neutrino/antineutrino-Gd
cross sections at neutrino energies below 80 MeV, considering the
most abundant even isotopes of Gadolinium that is, isotopes with
mass number A=156,158 and 160 (20.47\%, 24.84\% and 21.86\%
abundant, respectively). In this article, we have expanded this
calculation list and calculated the cross-sections of charged
current neutrinos and antineutrinos scattering off the stable odd
gadolinium isotopes, $^{155}$Gd and $^{157}$Gd (14.8\% and 15.6\%
abundant, respectively). The energy of impinging neutrinos is
related to supernova neutrinos while the corresponding nuclear
matrix elements have been calculated in the framework of
microscopic quasiparticle-phonon model.

\section{Brief description of the formalism}

The   standard model effective Hamiltonian in  the  charged
current reactions
\begin{eqnarray} \label{ch}
(A,Z)+ {\nu}_{e}\rightarrow (A,Z+1)+e^-\nonumber\\
(A,Z)+\bar{\nu}_{e}\rightarrow (A,Z-1)+e^+\nonumber
\end{eqnarray}
can be    written
\begin{equation} \label{hamil}
{\cal H} \, =\, \frac{G_F \hspace{3pt}cos\theta_c}{\sqrt{2}}
j_{\mu} ({\bf x}) J^{\mu} ({\bf x}),
 \end{equation}
A(Z) represents the mass(proton) number of a nucleus,
respectively. Here $G_F =1.1664\times 10^{-5}$$ GeV^{-2}$ denotes
the Fermi weak coupling constant and $\theta_c\simeq 13^o$ is the
Cabibbo angle. According to V-A theory, the leptonic current takes
the form \cite{Walecka,Donnelly,Donnelly1,Donnelly2}
\begin{equation}\label{lcurrent}
j_\mu = \bar{\psi}_{\nu_{\ell}}(x)\gamma_\mu (1 - \gamma_5)
 \psi_{\nu_{\ell}}(x)\, ,
\end{equation}
where $\psi_{\nu_{\ell}}$ are the neutrino/antineutrino spinors.
 The hadronic current of vector, axial-vector and
pseudo-scalar components  is written as
 \begin{eqnarray} \label{hcurrent}
J_{\mu}=\bar{\Psi}_N\big[ F_1 (q^2)\gamma_{\mu}+F_2 (q^2)\frac{i
\sigma_{\mu \nu}q^{\nu}}{2M_N}+F_A (q^2)\gamma_{\mu}\gamma_5 \nonumber\\
+F_P(q^2)\frac{1}{2M_N}q_{\mu}\gamma_5 \big]\Psi_{N}
\end{eqnarray}
($M_N$ stands for the nucleon mass, $\Psi_{N}$ denotes the nucleon
spinors and  $q^2$, the square of the four-momentum transfer). By
the conservation of the vector current (CVC), the vector form
factors  $F_{1,2}(q^2)$ can be written in terms of the proton and
neutron electromagnetic form factors \cite{athar06}.
 The axial-vector form factor $F_A(q^2)$ is
assumed to be of dipole form  \cite{Singh} while   the
pseudoscalar form factor $F_P(q^2)$ is obtained from the
Goldberger-Treiman relation~ \cite{Walecka}.

In the convention we used in the present work   the square of the
momentum transfer, is written as
\begin{equation} \label{qeq} q^2=q^{\mu}q_{\mu} =\omega^2- {\bf q^2} =
(\varepsilon_i - \varepsilon_f)^2-({\bf p}_i-{\bf p}_f)^2 \, ,
\end{equation}
where $\omega=\varepsilon_i-\varepsilon_f$ is the excitation
energy of the final nucleus. $\varepsilon_i$(${\bf p}_i$) denotes
the energy(3-momenta) of the incoming neutrino/antineutrino and
$\varepsilon_f$(${\bf p}_f$)  those of the outgoing
electron/positron,
 respectively.
The charged-current  neutrino/antineutrino-nucleus  cross section
is written as \cite{Donnelly}
\begin{eqnarray}
\label{eq:Sec2_1}
\sigma(\varepsilon_i)=& \dis{ \frac{2 G_F^2
cos^2\theta_c}{2J_i+1}} \sum_f
 {|{\bf p}}_f|\varepsilon_f\int_{-1}^{1}d(\cos{\theta})
 F(\varepsilon_f, Z_f)\nonumber\\
 &\times\big(\sum \limits_{J=0}^\infty \sigma_{CL}^J(\theta)
+
  \sum \limits_{J=1}^\infty \sigma_{T}^J(\theta) \big)
\end{eqnarray}
$\theta$  denotes the lepton scattering angle. The summations in
Eq. (\ref{eq:Sec2_1}) contain the contributions $\sigma_{CL}^J$,
for the Coulomb $\widehat{\mathcal{ M}}_J$ and longitudinal
$\widehat{\mathcal{ L}}_J $, and $\sigma_{T}^J$, for the
transverse electric $ \widehat{\mathcal{T}}_J^{el}$ and magnetic $
\widehat{\mathcal{T}}_J^{mag}$  multipole operators defined as in
Ref.~\cite{Divarijpg2}. These operators include both polar-vector
and axial-vector weak interaction components.

\begin{table}[htb]
 \caption{Adjusted (Adj) single-particle energies
  together with the Woods-Saxon (WS) energies   (in MeV) for
the neutron ($n$) and proton ($p$) orbitals.}
\begin{center}
\begin{tabular}{lclllllll}
\hline\Tstrut\Bstrut
orbital  & $^{154}$Gd    & &  $^{156}$Gd  &  & $^{158}$Gd  & \\
\hline
                &  WS   & Adj    & WS &  Adj  & WS &  Adj \\
\hline
n2p$_{1/2}$ &          &          &   -3.5015&   -4.8435&          &         \\
n2p$_{3/2}$ &   -4.5188&   -5.7000&   -4.4950&   -5.5000&   -4.4727&   -5.8000 \\
n1f$_{5/2}$ &           &          &   -3.6510&   -6.3407&   -3.6580&   -4.4000 \\
n1f$_{7/2}$ &   -6.4707&   -4.6800&   -6.4352&   -3.9263&   -6.4017&   -6.5000 \\
n0h$_{9/2}$ &           &         &   -5.6821&   -3.8911&   -5.7082&   -4.2000 \\
n0h$_{11/2}$&           &         &  -11.0539&   -1.4720&          &    \\
p1d$_{3/2}$ &   -3.9148&   -6.6000&   -4.4795&   -6.7000&   -5.0446&   -6.0000 \\
p1d$_{5/2}$ &   -6.2413&   -7.7000&          &    &          -7.3505&   -7.9000 \\
p0g$_{7/2}$ &          &          &          &    &          -8.1939&   -7.2000 \\
p0g$_{9/2}$ &          &          &          &    &         -12.7976&  -10.0000 \\
p0h$_{11/2}$&          &           &         &    &          -6.0117&  -12.0000 \\
 \hline
\end{tabular}
\end{center}
\label{adj}
\end{table}

\section{MQPM}

The microscopic quasiparticle-phonon model (MQPM ) is a
microscopic nuclear model which treats the structure of the odd-A
nuclei. The Hamiltonian can be written in the form
\begin{eqnarray}
\label{eqmqpm1}  H = \sum_{\alpha} E_a a^{\dagger}_{\alpha}
a_{\alpha} + H_{22} + H_{40}
           + H_{04} + H_{31} + H_{13}
\end{eqnarray}
were $E_{\alpha}$ are the quasiparticle energies  and other terms
of the hamiltonian are normal ordered parts of the residual
interaction labeled according to the number of quasiparticle
creation and annihilation operators which they contain.

The $H_{22}$,$H_{40}$ and $H_{04}$ parts of the quasiparticle
Hamiltonian are  treated in the BCS and quasiparticle random-phase
approximation (QRPA) framework and leads to definition of the
quasiparticles and the excitation (phonon) spectrum of the
doubly-even reference nucleus \cite{Ring-Book}. The $H_{31}$ and
$H_{13}$ parts are then diagonalized in the quasiparticle-phonon
basis and the eigenvectors represent the spectrum of the odd-A
nucleus.

The approximate ground state of the even-even reference nucleus is
obtained from a BCS calculation, where quasiparticle energies and
occupation factors $u_a$ and $v_a$ are obtained from the
Bogoliubov-Valatin transformation to quasiparticles

\begin{eqnarray}
\label{BV}
       a^{\dagger}_{\mu} &=& u_{\mu}c^{\dagger}_{\mu}-v_{\mu}\tilde{c}_{\mu}
       \nonumber \\
       \tilde{a}^{\dagger}_{\mu} &=& u_{\mu}\tilde{c}^{\dagger}_{\mu}
       +v_{\mu}c_{\mu}
\end{eqnarray}
where $\tilde{a}^{\dagger}_{\mu}=a^{\dagger}_{-\mu}(-1)^{j+m}$ and
$\tilde{c}^{\dagger}_{\mu}=c^{\dagger}_{-\mu}(-1)^{j+m}$. In Eq.
(\ref{BV}) $c^{\dagger}_{\mu}$ is the particle creation operator
and $\tilde{c}^{\dagger}_{\mu}$ denotes the time-reversed particle
annihilation operator.

  In
the QRPA the creation operator for an excited state (QRPA phonon)
has the form

\begin{equation}
      Q^{\dagger}_{\omega} = \sum_{a\le a'} \left\lbrack X^{\omega}_{aa'}
      A^{\dagger}(aa';J_{\omega}M) -
      Y^{\omega}_{aa'} {\tilde{A}}(aa';J_{\omega}M) \right\rbrack
\end{equation}
where the quasiparticle pair creation and annihilation operators
are defined as $A^{\dagger}(aa';JM)=\sigma^{-1}_{aa'}\Bigl\lbrack
a^{\dagger}_{a} a^{\dagger}_{a'} \Bigr\rbrack_{JM}$,
${\tilde{A}}(aa';JM)=\sigma^{-1}_{aa'}\Bigl\lbrack
\tilde{a}_{a}\tilde{a}_{a'} \Bigr\rbrack_{JM}$ and
$\sigma_{aa'}=\sqrt{1+\delta_{aa'}}$. Here the greek indices
$\omega$ denote phonon spin $J$ and parity $\pi$. Furthermore,
they contain an additional quantum number $k$ enumerating the
different QRPA roots for the same angular momentum and parity.
Thus $\omega=\lbrace J_{\omega},\pi_{\omega},k_{\omega} \rbrace$.

In the MQPM a state with angular momentum $j$ and projection $m$
in an odd-A nucleus is then created by using  the following
operator
\begin{equation}
\label{MQ}
     \Gamma_k^{\dagger}(jm)   =
      \sum_n D^k_n a^{\dagger}_{njm} + \sum_{\omega\alpha}
      D^k_{\alpha\omega} \Bigl\lbrack a^{\dagger}_{\alpha}
      Q^{\dagger}_\omega
      \Bigr\rbrack_{jm}
\end{equation}
The MQPM states with one dominant amplitude $D^k_n$ are called
one-quasiparticle-like and states with one,few or many important
amplitudes $D^k_{\alpha\omega}$ are called
three-quasiparticle-like states. For more details about the MQPM
see Ref~\cite{Suhonen-Book}.

\begin{figure}[htb]
\begin{center}
 \includegraphics[width=0.45\textwidth]{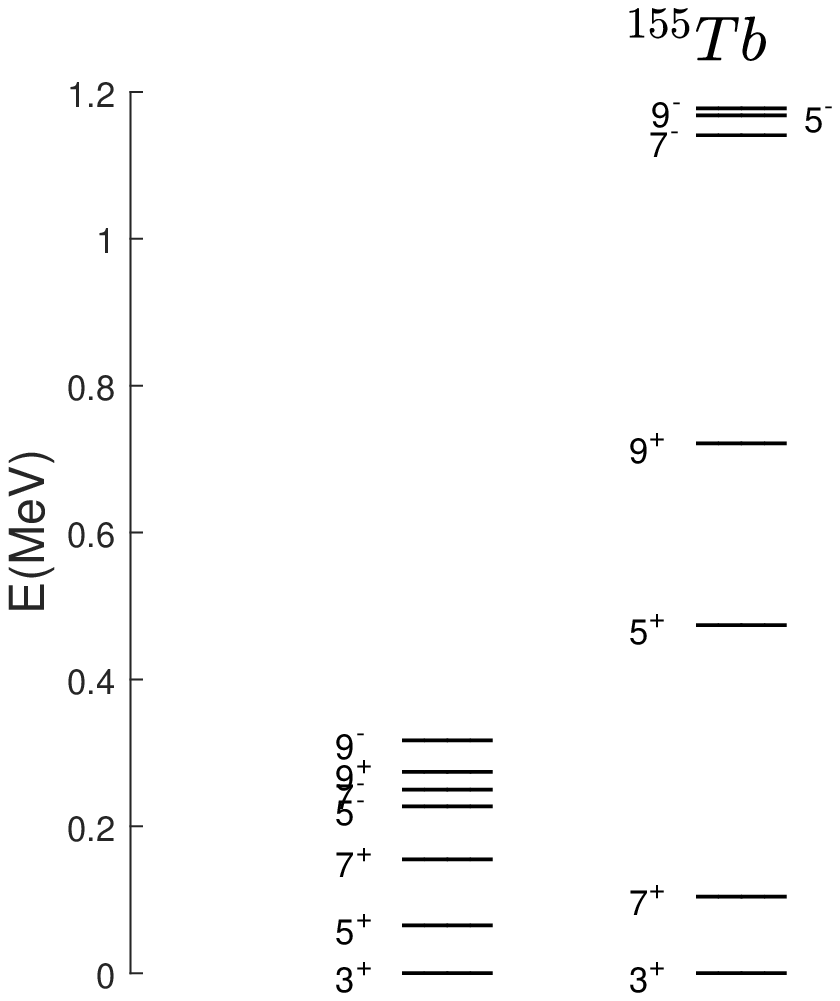}
 \includegraphics[width=0.45\textwidth]{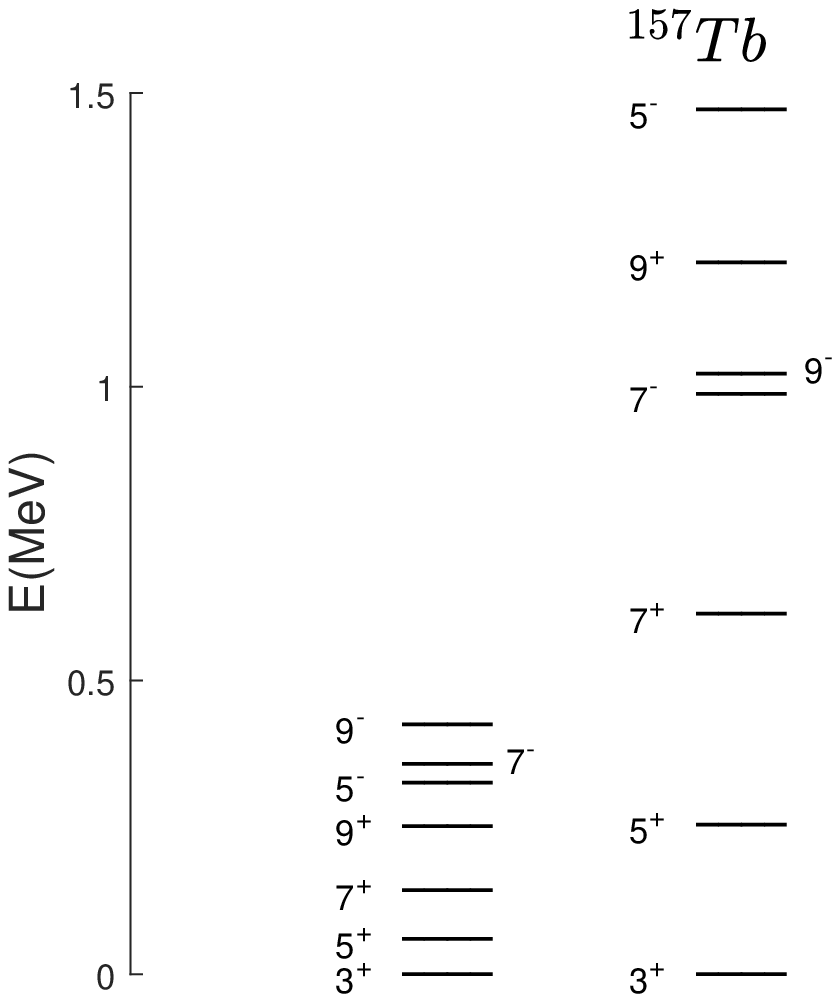}
\caption{(Color on line)  Experimental (left)\cite{tb155,tb157}
and theoretical (right) spectra of $^{155}$Tb   and   $^{157}$Tb}
\label{fasma}
\end{center}
\end{figure}

\section{Results}

For the reactions $^{A}Gd(\bar\nu_e , e^+)^{A}Eu$, A=155,157 the
initial sates of  $^{155}$Gd ( $^{157}$Gd) are taken as neutron
hole states of the reference nucleus $^{156}$Gd ( $^{158}$Gd)
while those of $^{155}$Eu ($^{157}$Eu) as proton hole states.
 On the other hand for the reactions
 $^{A}Gd(\nu_e , e^-)^{A}Tb$ the initial states of $^{155}$Gd (
 $^{157}$Gd) are taken as neutron states of the reference nucleus $^{154}$Gd (
 $^{156}$Gd), while those of $^{155}$Tb ($^{157}$Tb) as proton
 ones.

The wave functions of   initial and final states are computed by
using the microscopic quasiparticle phonon model (MQPM)
(\cite{Suhonen-Book}). The active model space for protons consists
of the complete oscillator shells $4\hbar\omega$ and
$5\hbar\omega$ while for neutrons the   oscillator shells are
$5\hbar\omega$ and $6\hbar\omega$. The corresponding single
particle energies (s.p.e) were produced by the well known Coulomb
corrected Woods-Saxon potential adopting the parameters of Bohr
and Mottelson \cite{Bohr-Mot}. The pairing interaction between the
nucleons can be adjusted by solving the corresponding BCS
equations \cite{Divari2018}.

In addition, adjustments of some single-particle energies, by
comparison with spectra of the neighboring odd nuclei, were done.
The adjustments made are shown in Table~\ref{adj}. The two-body
matrix elements were obtained from the Bonn one-boson-exchange
potential  applying the G-matrix techniques \cite{Holinde}. The
phonon basis contains all QRPA states having angular momentum $J^{
\pi}\le 6^{\pm}$ and excitation energy $E\le 10$ MeV. As an example
the calculated  energy spectra of low-lying states up 1.5 MeV for
$^{155,157}$Tb isotopes together with the experimental  ones
\cite{tb155,tb157} are presented in Fig.~\ref{fasma}.

\begin{figure}[htb]
\begin{center}
\hspace{1cm}\includegraphics[width=0.45\textwidth]{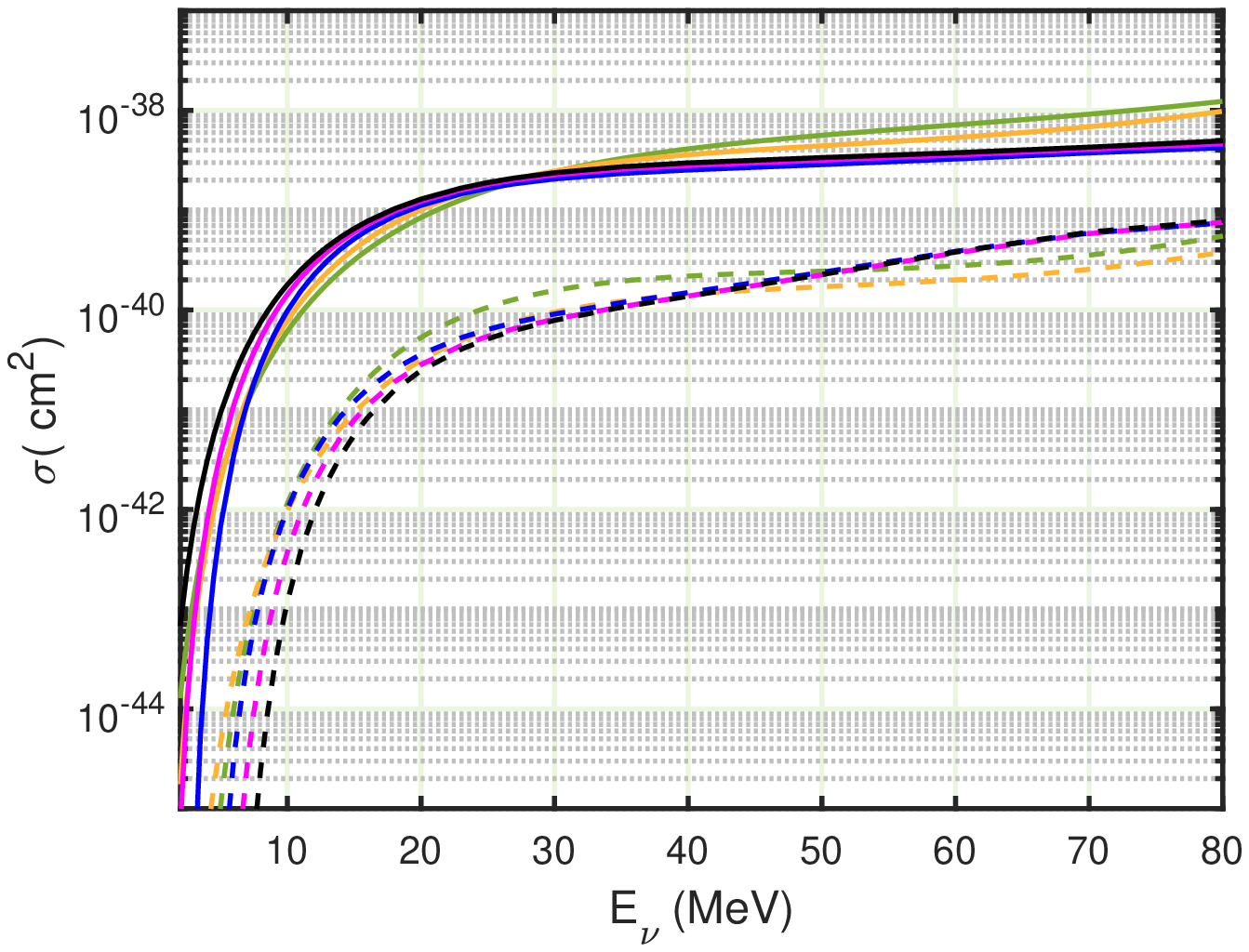}a)
\includegraphics[width=0.45\textwidth]{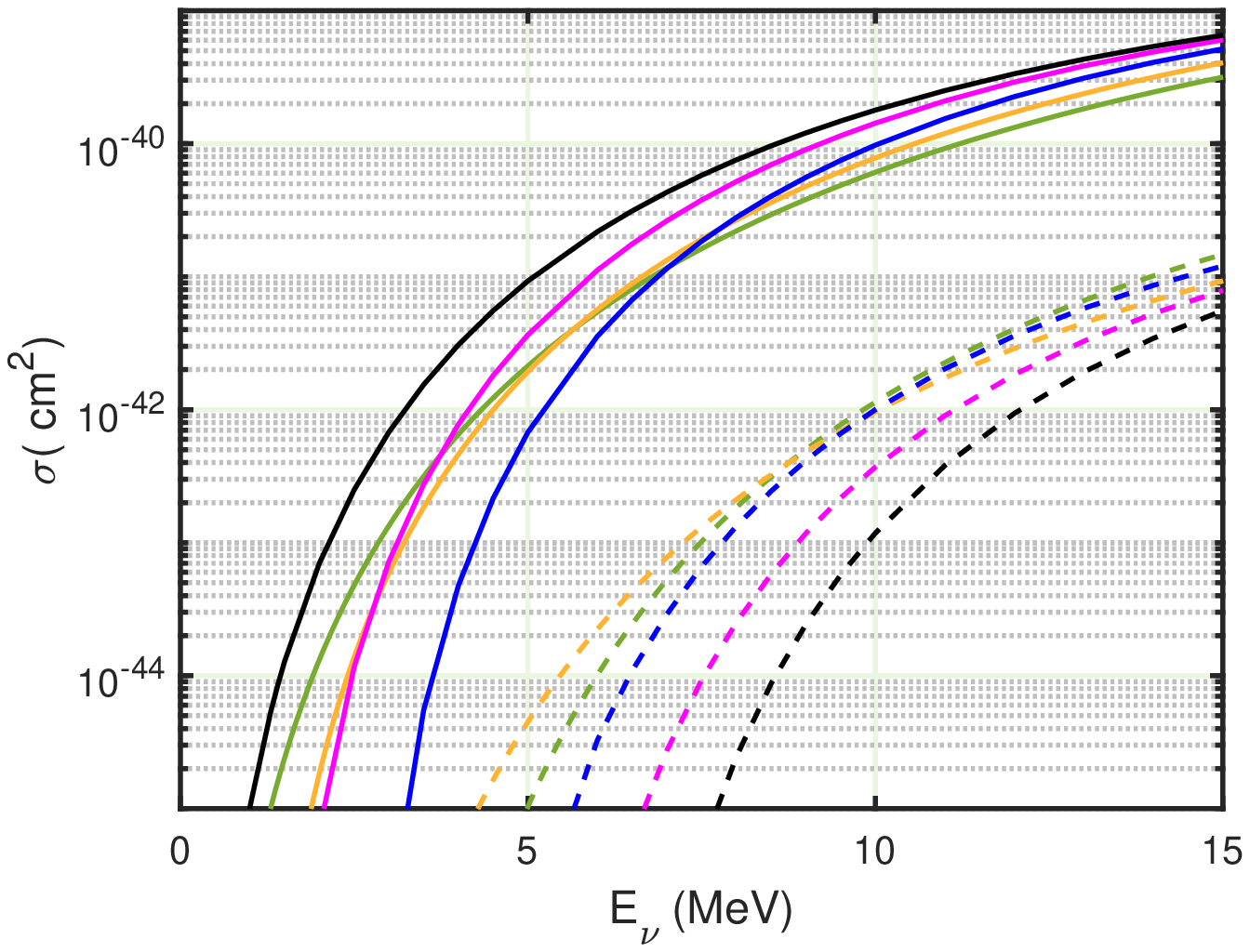}b)
\hspace{1cm}\includegraphics[width=0.45\textwidth]{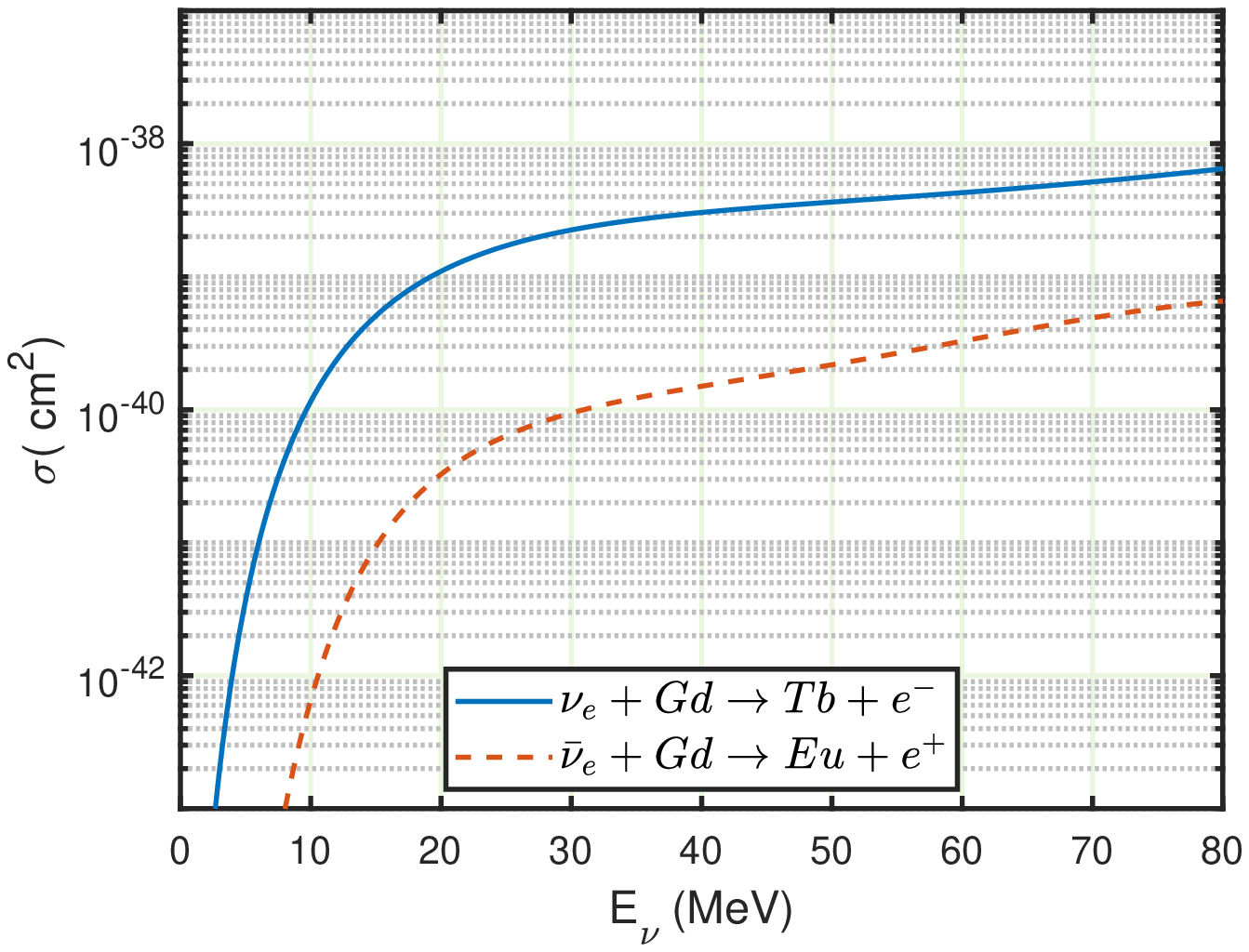}c)
\includegraphics[width=0.45\textwidth]{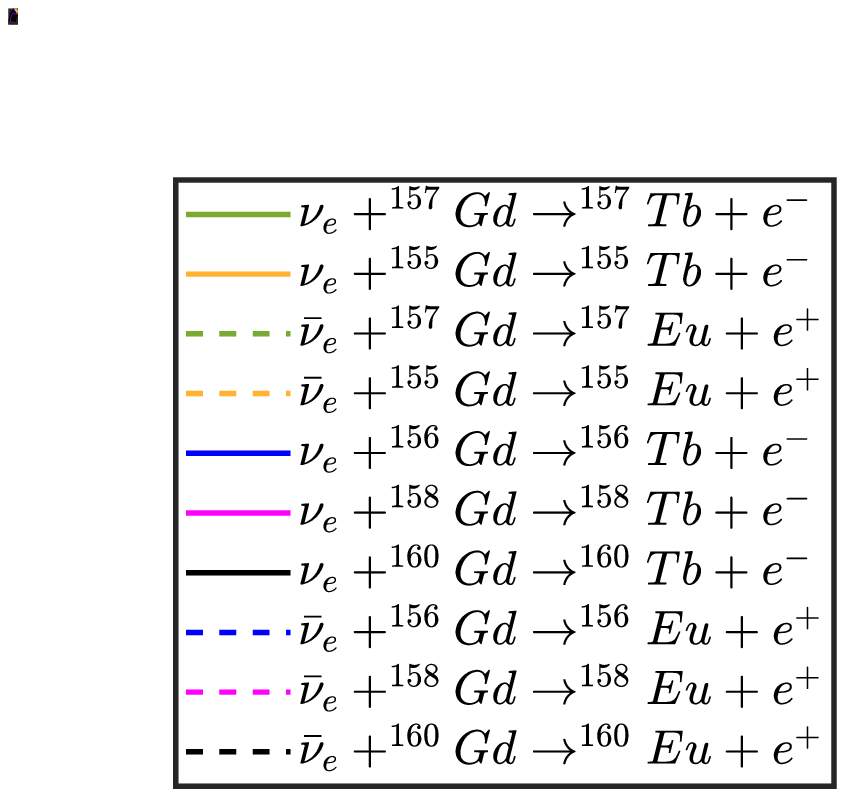}\hspace{-1cm}
 \caption{(Color on line)  (a) Total cross sections of  $\nu_e$ and
$\bar\nu_e$ Gd interactions for each isotope. (b) Total cross
sections of $\nu_e$ and $\bar\nu_e$ Gd interactions for each isotope
in the energy region from 0 to 15MeV. (c)Total cross sections of
natural   Gadolinium $Gd(\nu_e , e^-)Tb$ (solid line) and
$Gd(\bar{\nu}_e , e^+)Eu$ (dashed line)    taking into account the
corresponding abundances of each isotope. } \label{cs}
\end{center}
\end{figure}

\begin{figure}[htb]
\begin{center}
 \includegraphics[width=0.45\textwidth]{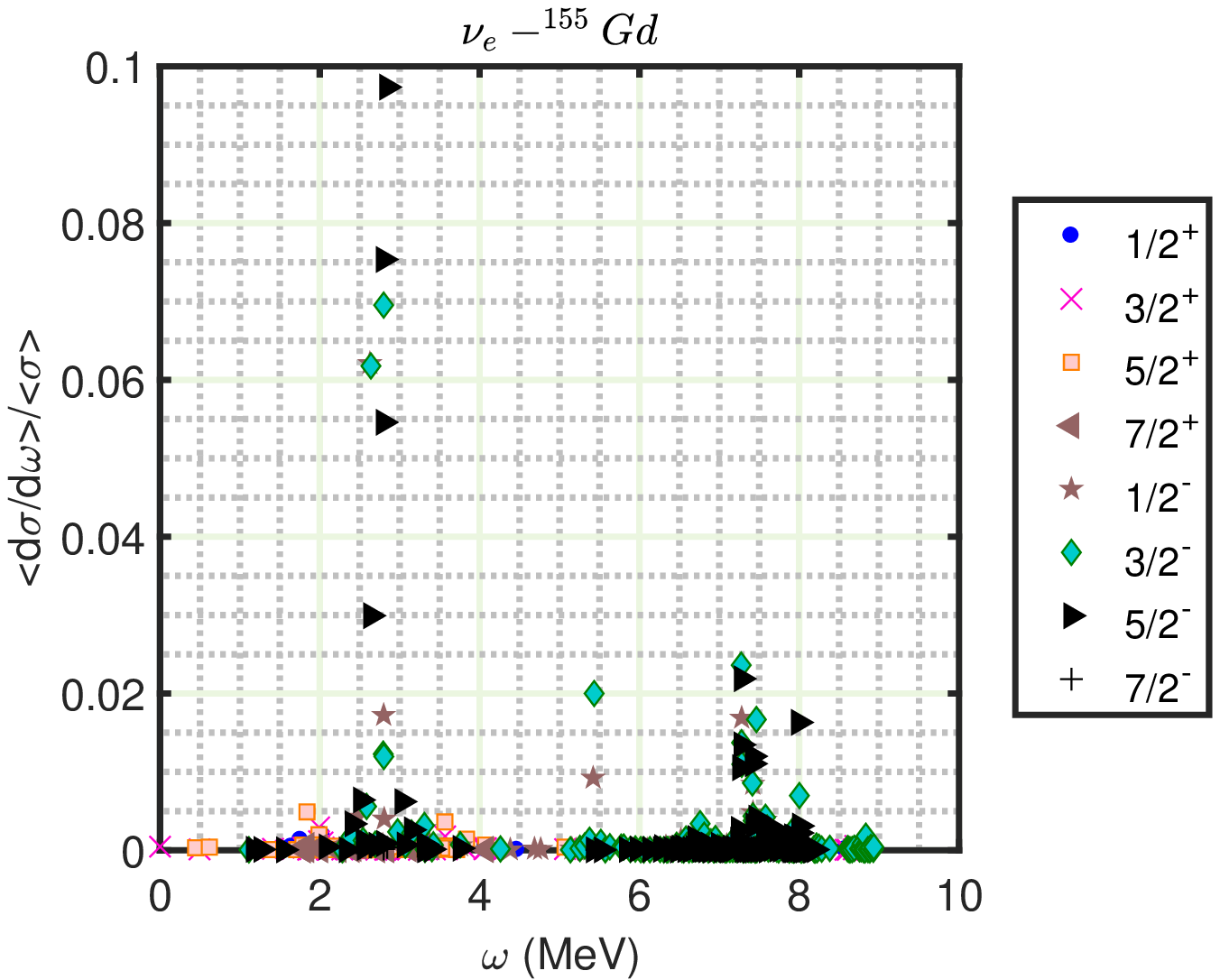}a)
 \includegraphics[width=0.45\textwidth]{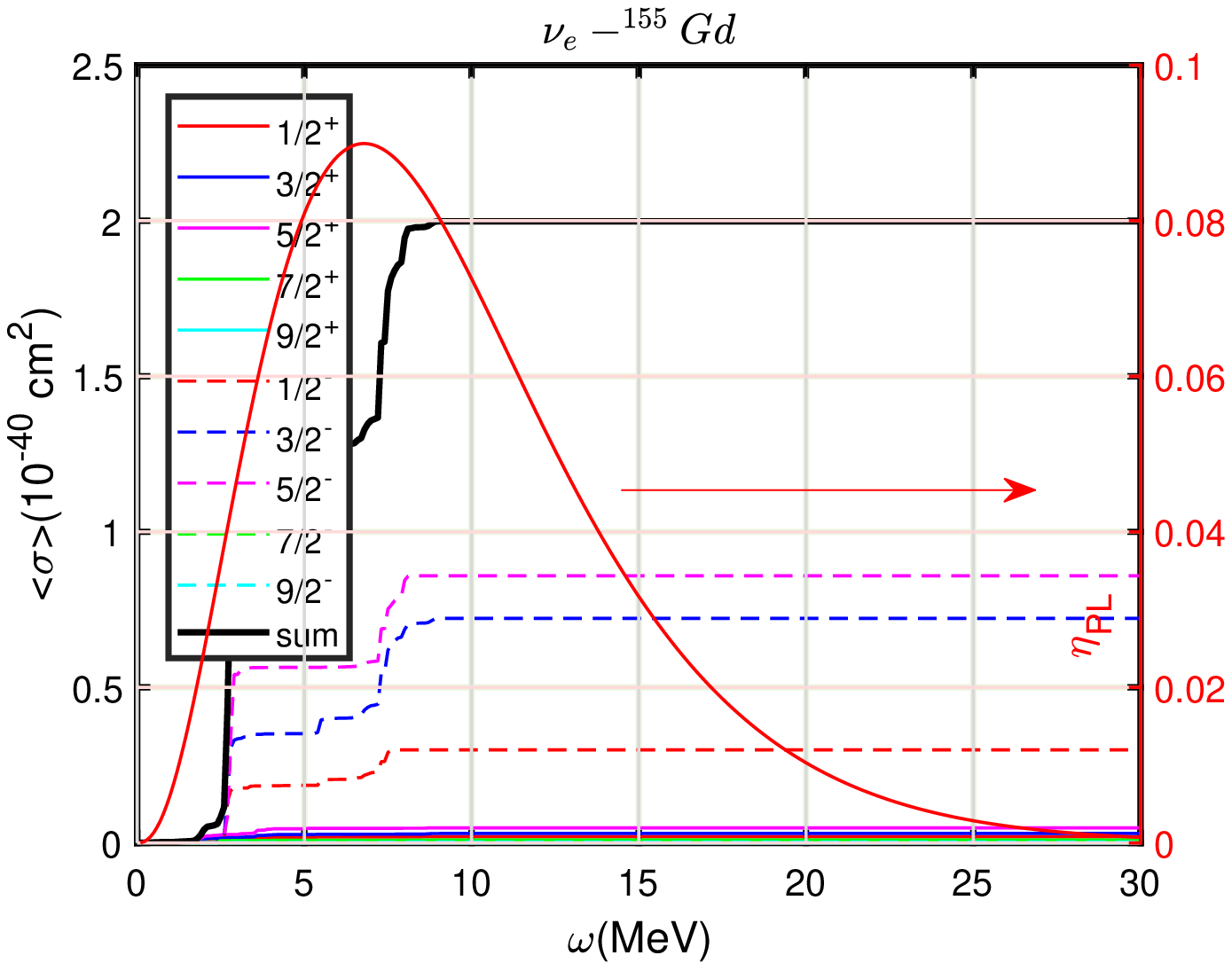}b)
 \includegraphics[width=0.45\textwidth]{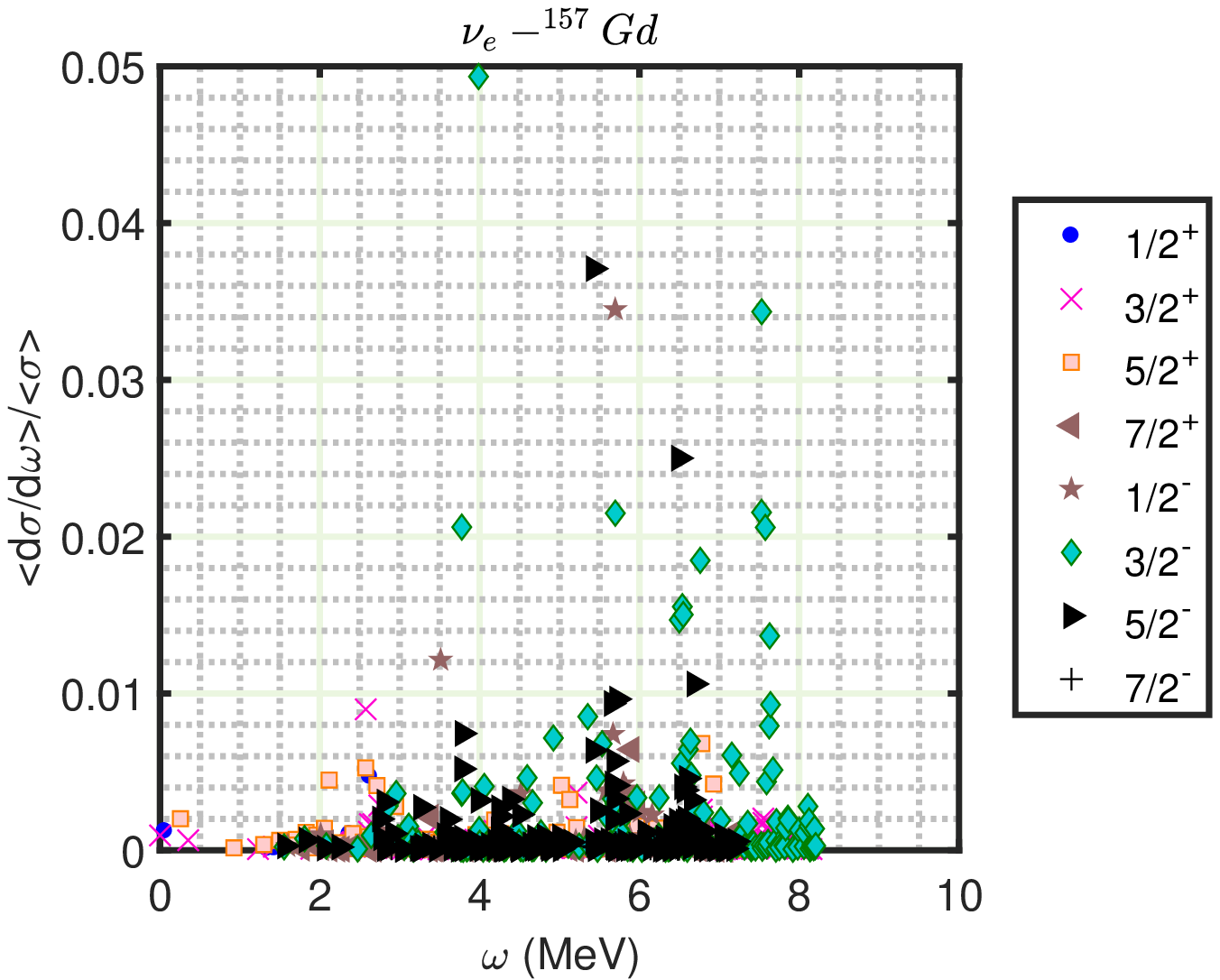}c)
 \includegraphics[width=0.45\textwidth]{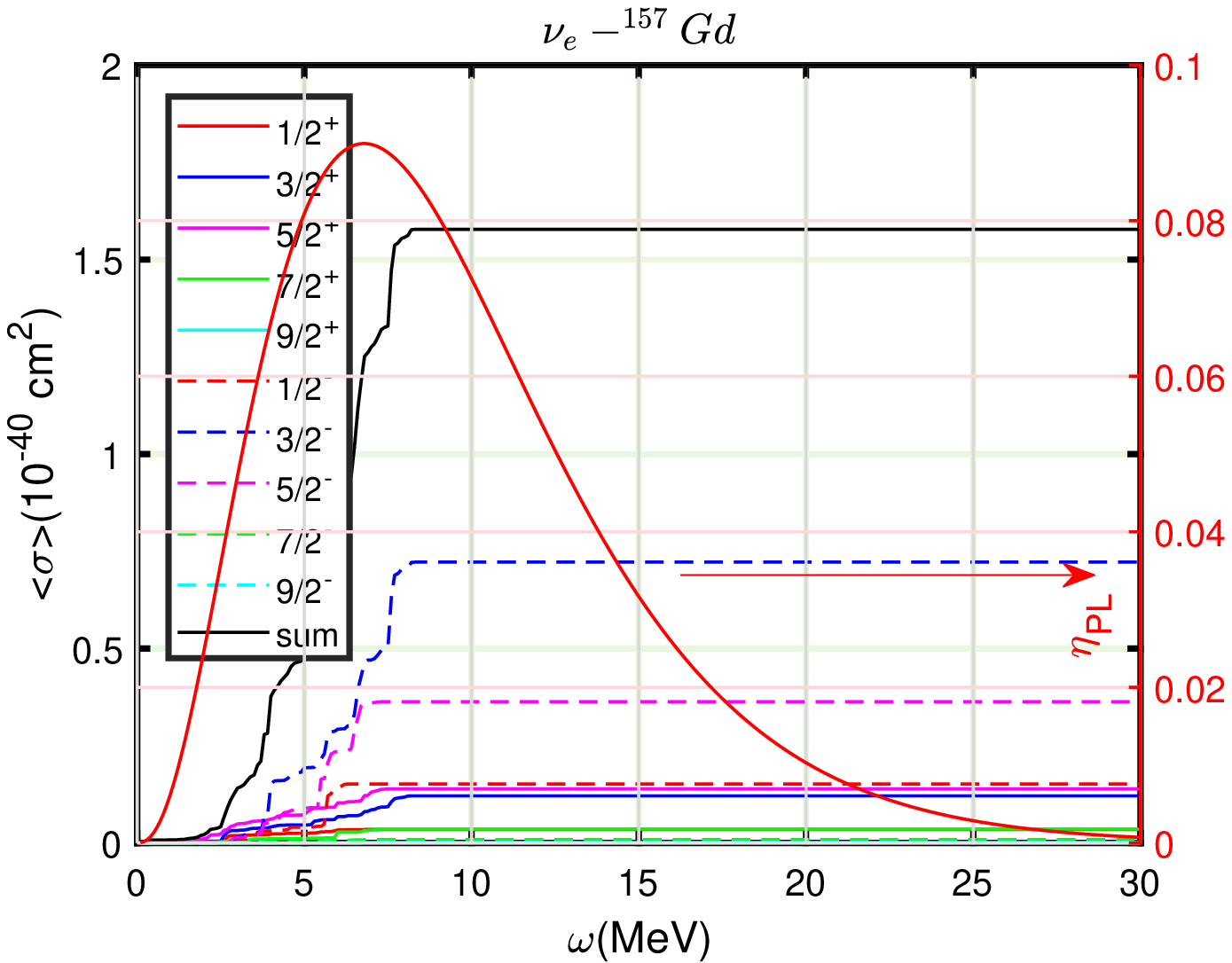}d)
 \caption{(Color on line) (a) and (c)  Differential cross sections (normalized) for the charged current neutrino
 interactions off  $^{155,157}$Gd to final states in  $^{155,157}$Tb. (b) and (d) Cumulative flux-averaged
cross sections (in units $ 10^{-40}cm^2$) as a function of
excitation energy $\omega$ for the reactions $^{A}Gd(\nu_e ,
e^-)^{A}Tb$, A=155,157. The power-law(PL)  distributions
$\eta_{PL}$ (red solid line) for $\langle E_{\nu_e}\rangle=9.5$MeV
and  $\alpha=2.5$ is also displayed. } \label{normTb}
\end{center}
\end{figure}

\begin{figure}[htb]
\begin{center}
 \includegraphics[width=0.45\textwidth]{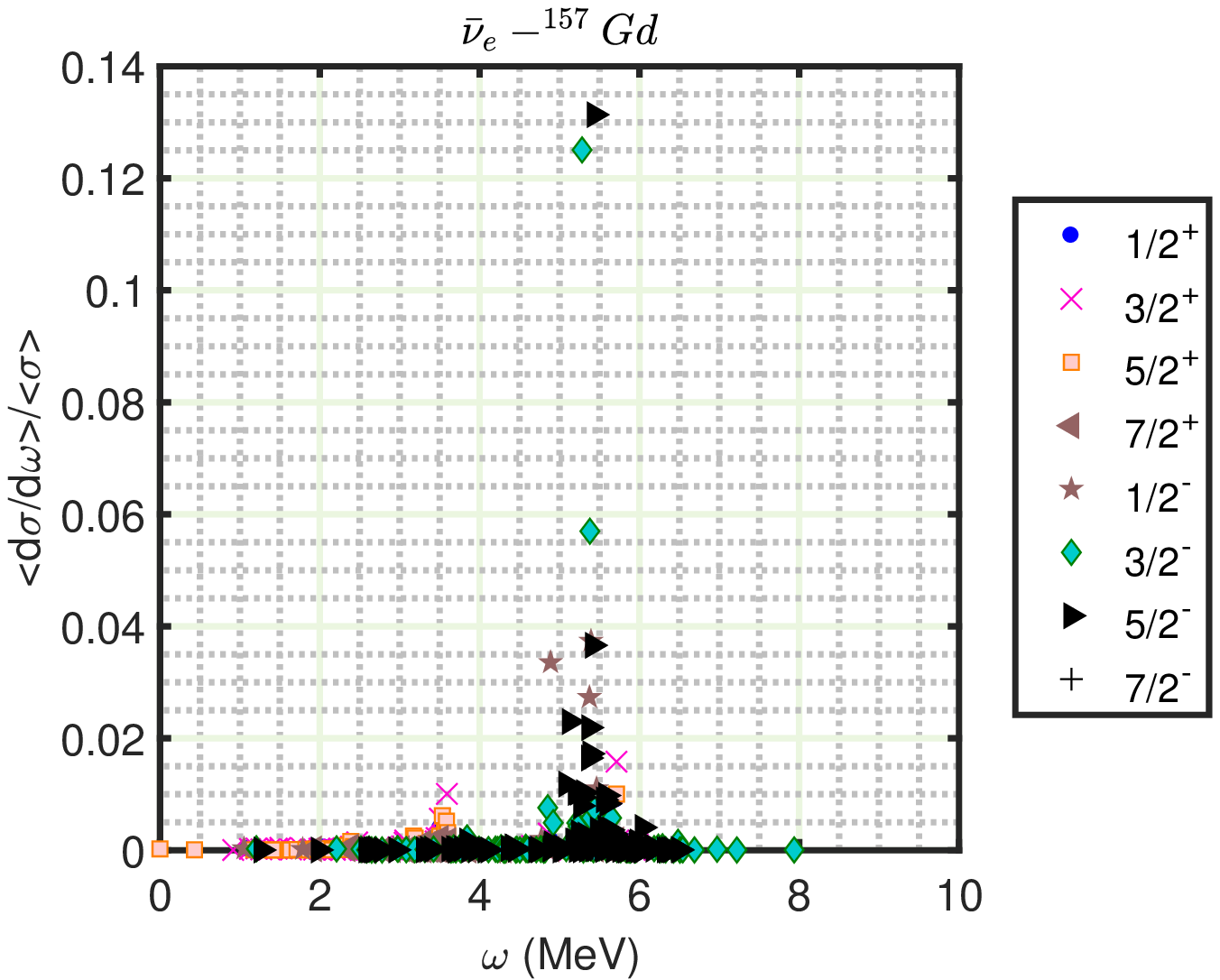}a)
 \includegraphics[width=0.45\textwidth]{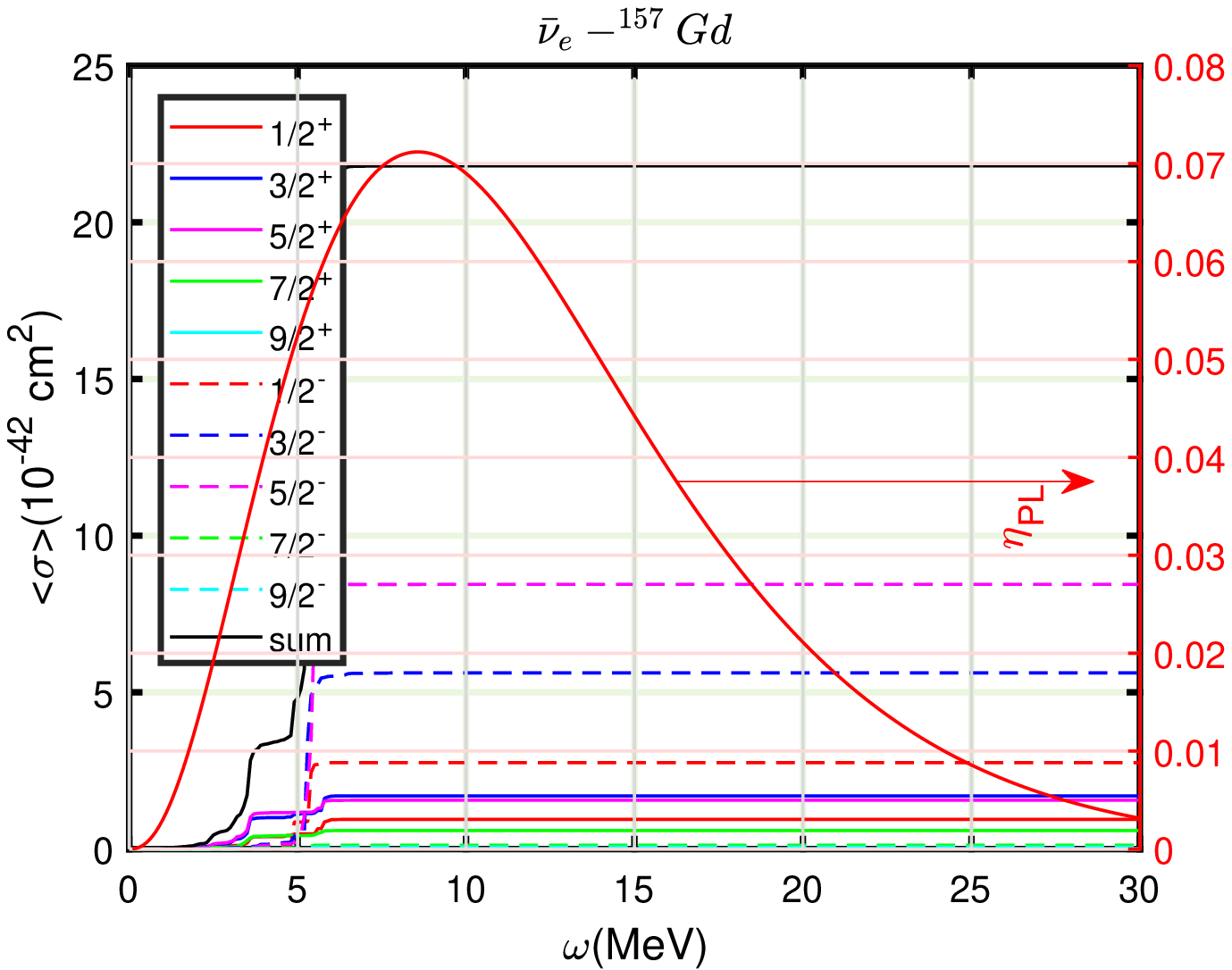}b)
 \includegraphics[width=0.45\textwidth]{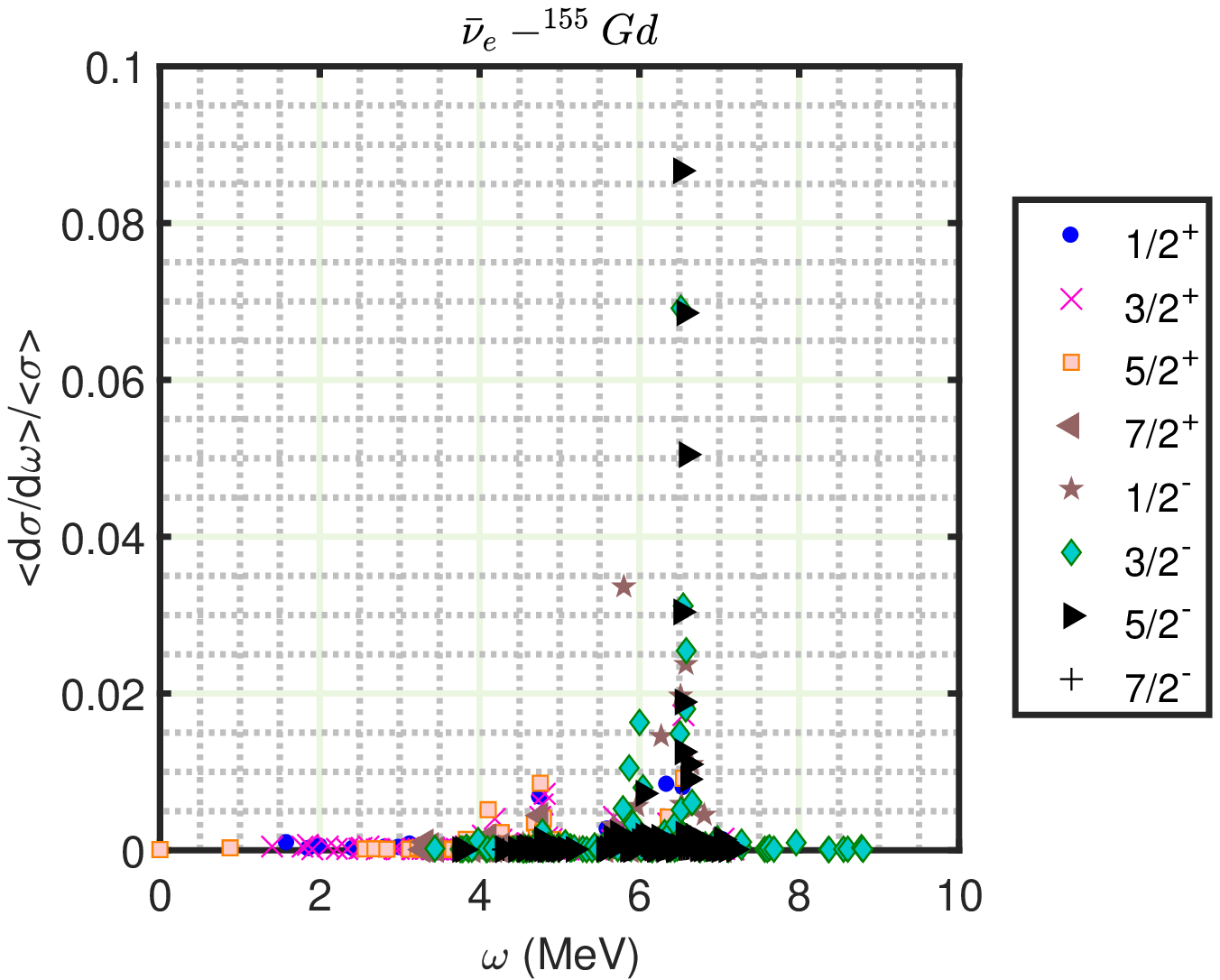}c)
 \includegraphics[width=0.45\textwidth]{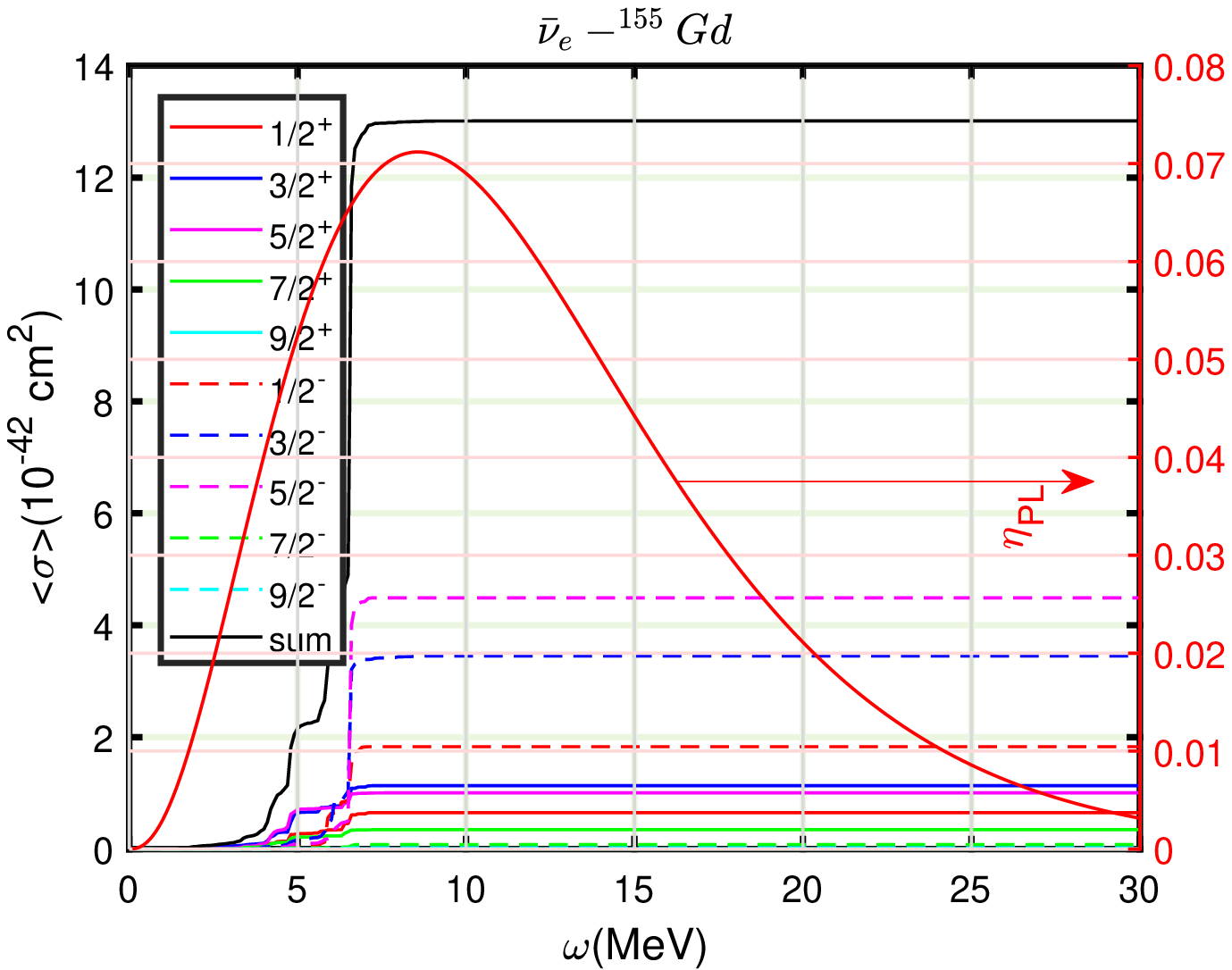}d)
 \caption{(Color on line) (a) and (c)  Differential cross sections (normalized) for the charged current neutrino
 interactions off  $^{155,157}$Gd to final states in  $^{155,157}$Eu. (b) and (d) Cumulative flux-averaged
cross sections (in units $ 10^{-42}cm^2$) as a function of
excitation energy $\omega$ for the reactions $^{A}Gd(\nu_e ,
e^-)^{A}Eu$, A=155,157. The power-law(PL)  distributions
$\eta_{PL}$ (red solid line) for $\langle E_{\nu_e}\rangle=12$MeV
with $\alpha=2.5$ is also displayed.   } \label{normEu}
\end{center}
\end{figure}

In Fig.~\ref{cs} (a) and (b) we display the numerical results of
the total scattering cross section $\sigma(E_{\nu})$ given by
Eq.~(\ref{eq:Sec2_1})  as a function of the incoming neutrino
energy $E_{\nu}$ for the reactions $^{A}Gd(\nu_e , e^-)^{A}Tb$ and
$^{A}Gd(\bar{\nu}_e , e^+)^{A }Eu$, A=155,157 respectively, along
with the results concerning the even isotopes A=156,158,160 taken
from Ref.~\cite{Divari2018}.
In Fig.~\ref{cs} also shown are the total cross sections of
 $Gd(\nu_e , e^-)Tb$ (solid line) and
$Gd(\bar{\nu}_e , e^+)Eu$ (dashed line) for natural Gadolinium(the
natural abundance of each isotope has been taken into account).
The normalized differential cross sections for the charged current
neutrino(anti-neutrino) scattering leading to final states are
shown in Figures  \ref{normTb}(a),(c) and \ref{normEu}(a),(c) for
the iotopes $^{155,157}$Tb($^{155,157}$Eu), respectively.
It can be  seen that  the dominant transition is to the final
states ${1/2}^-$, ${3/2}^-$ and ${5/2}^-$. In addition, the final
nuclear state distribution in the neutrino-induced reactions is
within a wide range of excitation energy 2 to 8 MeV. In contrast,
for antineutrino scattering, the most prominent transitions are
lying between 5 to 7 MeV.


The number of expected neutrino events is estimated in a WCD using
our predictions of the total   cross sections   taking
  0.1\% (by mass) of Gd doping \cite{Beacom:2003nk}.
In the SK detector where the fiducial mass of water is 32~ktons
the Gadolinium mass there would be about $m_t=32$~tons. Supernova
radiates $3\times10^{53}$ erg of total energy in about $10~s$
through neutrinos. Assuming equal energy partition among
neutrinos, supernova radiates $N_{\nu_e}=3.0\times 10^{57}$
electron neutrinos and $N_{\bar\nu_e}=2.6\times 10^{57}$ electron
antineutrinos.
 The fluence $\Phi(E_{\nu})$ for neutrinos integrated over 15
s burst is given by the relation
\begin{equation}
\label{fluxeq} \Phi_i(E_{\nu})=\frac{N_i}{4\pi D^2}
 \eta_{PL}(E_{\nu}), \quad i=\nu_e,\bar\nu_e
\end{equation}
at a distance $D=10$~kpc=$3.1\times 10^{22}$cm. Where
$\eta_{PL}(E_\nu)$ being the  power-law distribution
\cite{Lujan-Peschard:2014lta,Tamborra:2012ac}:
\begin{eqnarray}\label{pldis}
\eta_{PL}(E_\nu)=
    \frac{E_{\nu}^{\alpha}\:
    e^{-E_{\nu}/T_i}}{T_i^{\alpha +1 }\:
    \Gamma\left(\alpha+1\right)}
\end{eqnarray}
with
\begin{equation}
    T_i = \frac{\langle E_i\rangle}{(\alpha + 1)}, \quad i = \nu_{\mathrm{e}},\,
    \overline{\nu}_{\mathrm{e}}
    \label{eq:TEmedia}
\end{equation}
 If the mass of the target material is $m_t$, corresponding
to $N_{t}$ atoms then the number of expected events   are
\begin{equation}
 N_{event}   =N_{t}\int \Phi_i(E_{\nu})\sigma_i(E_{\nu})dE_{\nu}=
 N_{t}\frac{N_i}{4\pi D^2}\langle \sigma_i\rangle \label{exam1}
\end{equation}
where $\langle \sigma_i\rangle$ being the flux-averaged cross
section.

\begin{figure}[htb]
\begin{center}
\includegraphics[scale=0.45]{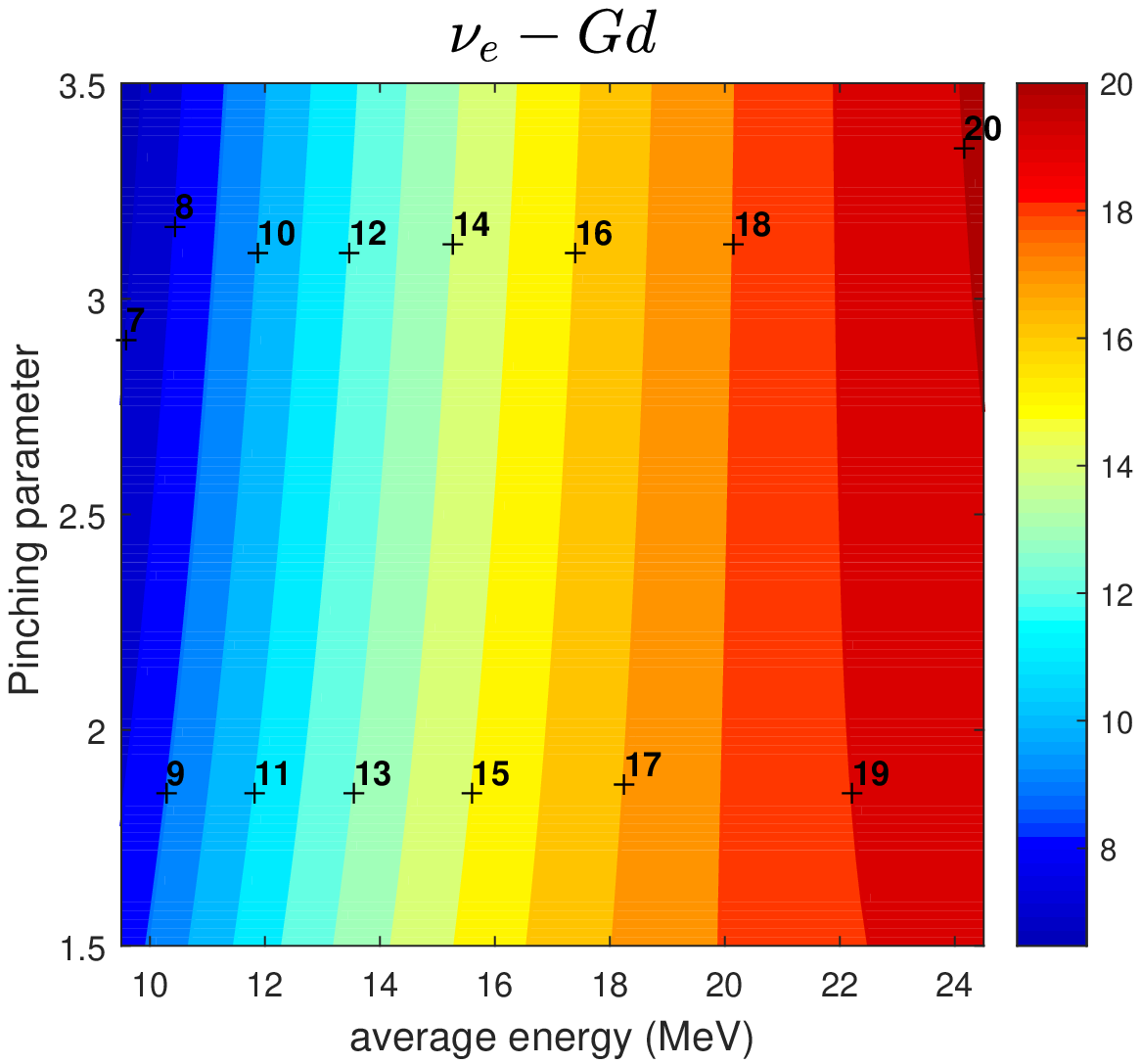}
\includegraphics[scale=0.45]{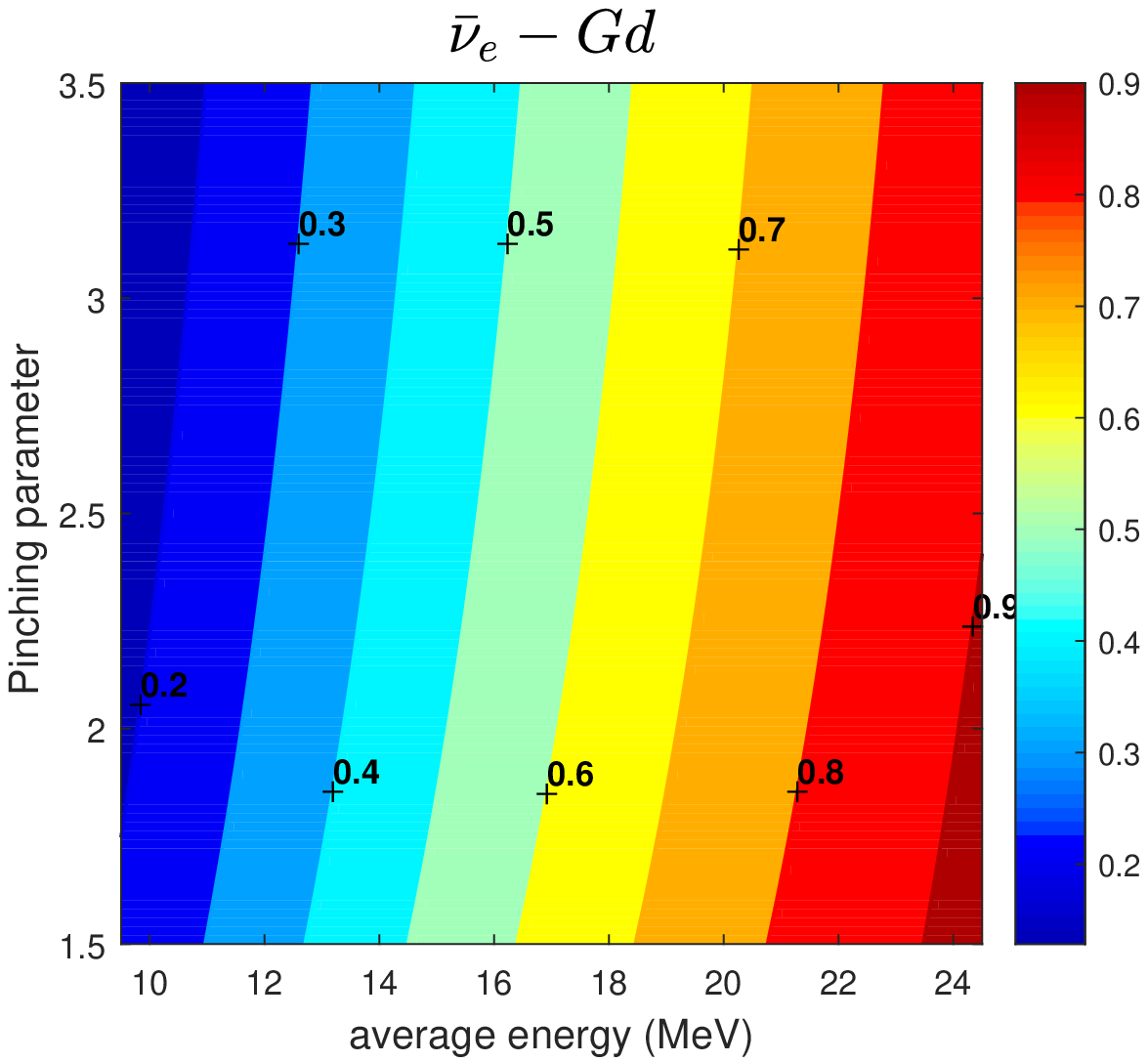}
\caption{(Color on line) These contour plots show the number of
electrons (left panels) and positrons (right panels) emitted from
32~tons of natural Gd. The contours from left to right in each panel
indicates the increase in the number of expected events.}
\label{alphaT}
\end{center}
\end{figure}

Neutrinos in water can be   detectable through the channels
$\bar\nu_e+p$ (IBD), the elastic scattering $\nu_e+e^-$ (ES) as well
as the neutral current scattering $\nu_e+^{16}O$ (OS) on oxygen.  As
it is known
the largest number of events will be due to the IBD which is almost
isotropic \cite{Vogel:1999zy}. Neutrons produced in the IBD   are
captured by Gd nuclei. These neutrons are easy to see because the
Cherenkov light generated by the electrons was struck by a gamma
cascade of about 8 MeV emitted by the Gd nucleus.
In Fig.~\ref{alphaT} a contour plot is used to display the number
of expected events for the reactions $Gd(\nu_e,e^-)Tb$ and
$Gd(\bar\nu_e,e^+)Eu$ respectively, with various parameterizations
of power-law spectra. As shown in the figure, in the window of
10-18 MeV, the number of events almost depends on the pinching
parameter $\alpha$. Furthermore, the $\bar\nu_e-$Gd events are
quite small and are hidden by the large IBD interactions on free
protons \cite{Divari2018}. The ability to completely isolate major
IBD events is extremely important for studying the remaining
reactions $\nu_e-$Gd that cause the gamma-ray emission.
If $\nu_e-$Gd   events can be isolated by gamma-ray identification
or by determining possible delayed beta decay, then they may have
certain advantages due to low thresholds (although low yields).
Recently it is shown that the Gd-ion release from a custom designed
glass in the form of beads or powders   could be used as a
controlled Gd-source in future WCDs to enhance neutrino detection
\cite{Dongol}.

The   flux-averaged supernova-neutrino ($SN-\nu$) cross sections,
for the various Gd isotopes, appear in  Table \ref{tab:fluxavernu}.
The pinching parameter $\alpha$ has been taken the value $
\alpha=2.5$, at a typical supernova neutrino (antineutrino) mean
energy $\langle E_{\nu_e}\rangle=9.5$~MeV ($\langle
E_{\bar\nu_e}\rangle=12$~MeV). It can be seen that the cross
sections of odd mass isotopes are of the same order of magnitude  as
the cross sections of even mass isotopes \cite{Divari2018}.
  The  cross sections for natural Gd in the reactions $Gd(\nu_e , e^-)Tb$
($205~10^{-42}cm^2$ ) and $Gd(\bar{\nu}_e , e^+)Eu$
($12.6~10^{-42}cm^2$),   have been  taken from   the corresponding
abundances of each isotope.
The results refer to an ideal detector operating down to zero
threshold $E_{th}=0$. In the case of non zero threshold the flux
 averaged cross sections it is expected to be  suppressed   by about 9\% for   an
electron total energy threshold of 5 MeV (energy threshold in
SK)Ref.~\cite{Divari2018}.

\begin{table}[tbp]
 \caption{ The first column gives the   neutrino
nucleus reaction and the second column gives the corresponding
mean energy $\langle E_{\nu}\rangle$. The third column gives the
total flux-averaged cross sections in units of $10^{-42}~cm^2$.
The pinching parameter is taken to be $\alpha=2.5$.}
\label{tab:fluxavernu}
\renewcommand{\tabcolsep}{0.3pc} 
\renewcommand{\arraystretch}{1.0} 
\begin{center}
\begin{tabular}{|c|c|c|c|c|c|c|c|c|c|c|}
\hline\Tstrut\Bstrut
                     & $\langle E_{\nu}\rangle$(MeV) &   $\langle \protect\sigma \rangle_{}~($$10^{-42}~cm^2)$   \\
\hline\hline\Tstrut\Bstrut
$\nu_e-^{155}$Gd     &9.5     & 199   \\
$\nu_e-^{156}$Gd     &9.5     & 200  \\
$\nu_e-^{157}$Gd     &9.5     & 158  \\
$\nu_e-^{158}$Gd     & 9.5    & 224  \\
$\nu_e-^{160}$Gd     & 9.5    & 241    \\
\hline\Tstrut\Bstrut
 $\nu_e-$Gd             & 9.5    & 205       \\
\hline \hline\Tstrut\Bstrut
$\bar\nu_e-^{155}$Gd &12      & 13.0  \\
$\bar\nu_e-^{156}$Gd &12      & 13.7  \\
$\bar\nu_e-^{157}$Gd &12      & 21.8   \\
$\bar\nu_e-^{158}$Gd &12      & 9.7   \\
$\bar\nu_e-^{160}$Gd &12      & 8.4  \\
\hline\Tstrut\Bstrut
 $\bar\nu_e-$Gd        &12      & 12.6     \\
\hline
\end{tabular}
\end{center}
\end{table}

\section{Conclusions}

In this article, we have expanded the   calculation of
Ref.~\cite{Divari2018}   to include  the cross-sections of charged
current neutrinos and antineutrinos scattering off the odd
gadolinium isotopes, $^{155}$Gd and $^{157}$Gd.  The neutrino
induced transitions to excited nuclear states are computed in the
framework of  quasiparticle-phonon model. We give the functional
relationship between the total cross section and the energy of the
neutrino (antineutrino), so that it is possible to obtain the
total cross section of the required neutrino energy distribution
in the range of 5 to 80 MeV.  The nuclear responses to supernova
neutrinos for the   nuclei under consideration have been
calculated by folding the cross sections with a power-law
quasi-thermal neutrino distribution. The odd mass isotope cross
sections we obtained are of the same order of magnitude as the
even mass isotope cross sections. The dominant transitions are to
the final states ${1/2}^-$, ${3/2}^-$ and ${5/2}^-$. The final
nuclear states in the neutrino-induced reaction are distributed in
a wide excitation energy range of 2 to 8 MeV, while for
antineutrino scattering, the most prominent transitions lie
between 5 and 7 MeV. The ability to completely isolate the major
IBD events is extremely important for studying the remaining
reactions $\nu_e-$Gd that cause   gamma-ray emission. If
$\nu_e-$Gd events can be isolated, then they may have certain
advantages due to low thresholds (although the yield is low).

The observation of the next (extra) galactic supernova is very
interested in astrophysics and particle physics. Although the
uncertainty of neutrino flux is still large, this observation will
bring important information about the explosion mechanism of the
neutrino light curve, as well as important information about
flavor conversion and energy spectrum reconstruction of supernova
neutrinos. Megaton-scale water detectors with GdCl$_3$ will have a
huge impact on the detection of supernova $\nu_e$. They can
observe hundreds of diffuse supernova neutrino background events
per year, so that they can strictly test the black hole formation
rate and supernova neutrino spectrum.

 \clearpage

\bibliography{TEX}
\end{document}